\documentstyle[aps,preprint]{revtex}
\begin{document}
\draft
\title{Chaos and fractals around black holes}
\author{C. P. Dettmann and N. E. Frankel}
\address{School of physics, University of Melbourne,
Parkville Victoria 3052, Australia}
\author{N. J. Cornish}
\address{Department of Physics, University of Toronto,
Toronto, Ontario M5S1A7, Canada}
\date{\today}
\maketitle
\begin{abstract}
Fractal basin boundaries provide an important means of
characterizing chaotic systems.  We apply these ideas to
general relativity, where other properties such as Lyapunov exponents are
difficult to define in an observer independent manner.  Here we discuss the
difficulties in describing chaotic systems in general relativity and
investigate the motion of particles in two- and three-black-hole spacetimes.
We show that the dynamics is chaotic by exhibiting the basins
of attraction of the black holes which have fractal boundaries.  Overcoming
problems of principle as well as numerical difficulties, we evaluate
Lyapunov exponents numerically and find that some trajectories have a positive
exponent.
\end{abstract}
\section{Introduction}

As we look out into the universe with ever more powerful telescopes, we
begin to see that the structures formed through gravitational organization
often have complicated, and sometimes fractal, geometry. The Voyager misson
revealed Saturn's rings to have an intricate distribution of rings
and voids, providing a fractal image of the underlying chaotic gravitational
dynamics. A similarly produced series of gaps can be seen in the asteroid belt
which lies between Mars and Jupiter~\cite{Wisdom1,Wisdom2}. Looking beyond our
solar system, the
large scale structure of the universe appears to be fractal over a range of
scales~\cite{Carruthers}, and
is commonly thought to be the result of chaotic evolution from smooth initial
conditions. Furthermore, when describing the universe using Einstein's theory
of gravity, space and time become dynamical and the evolution
of spacetime itself can be chaotic due to the non-linear nature of Einstein's
equations.

In this paper we study test particle trajectories in multi-black-hole
spacetimes.
We shall see that the phase space is divided into basins of attraction
which are separated by a fractal boundary. The presence of fractal boundaries
is due to the chaotic nature of the dynamics. It is no surprise that particle
trajectories can be chaotic in general relativity since many-body Newtonian
systems are known to be chaotic and Einstein's theory recovers Newtonian
gravity for weak fields and small velocities. What is more surprising is the
occurrence of chaotic attractors in a Hamiltonian system, a feature that is
foreign to non-relativistic celestial dynamics.
Attractors can arise in relativistic systems since there is a finite
maximum velocity, the speed of light, and even light can be captured by
a strong gravitational field. In the multi-black hole spacetime we shall
study, the black holes act as attractors in phase space. In fact, the whole
concept of what we mean by a phase space is fundamentally changed once
we consider general relativity. In the Newtonian view of the universe
space and time exists as a rigid underlying structure, providing a fixed
reference frame with which positions and velocities can be unambiguously
defined. In contrast, space and time are dynamical concepts in general
relativity and the coordinate system we use to describe notions of time and
place lose any fundamental significance.  Thus it is important to describe
the chaotic dynamics in terms which do not depend on the chosen coordinate
system.

The outline of this paper is as follows. In section II we discuss the subtle
issues that must be confronted when attempting to quantify chaos and fractals
in general relativity.
In the following sections, we investigate the motion of neutral and
charged test particles in
the combined gravitational and electric fields of two or three-black-holes,
following on from Refs.~\cite{C90,C91,CP,DFC}.  Each black hole
has charge equal to its mass, so that the total force between them is zero,
and only the motion of the particle need be considered.  This paper is the
first major study of the fractal boundary basin and Lyapunov exponents of
this system.  In section III we
write down the metric corresponding to this situation, noting in passing that
it is possible to write down exact solutions of the Einstein-Maxwell equations
containing fractal singularities.  The Hamilton-Jacobi method is used to show
that the two-black-hole problem is integrable in the weak field limit.
In section IV we look at qualitative features of this system which are
characteristic of chaotic dynamics, such as sensitive dependence on initial
conditions, complicated basin boundaries, and universality.  The effect of
changing the charge/mass ratio of the test particle and the mass of the
black holes is considered, and the three-black-hole problem is also treated.
Section V investigates quantitative indicators of chaos, that is, fractal
dimensions and Lyapunov exponents.

\section{Chaos and Fractals in General Relativity}
Studies of chaos in general relativity fall broadly into two categories.
Those in the first group look for chaos in the Einstein field equations.
After a coordinate system has been chosen
there are six nonlinear second order coupled partial differential equations,
so drastic simplifications are used to make analytic or numerical calculations
tractable.  The system receiving the most attention from the point of view of
chaos has been the Mixmaster Universe~\cite{Mis,BLK}.  This
is a general cosmological
model which is homogeneous but not isotropic.  Einstein's equations in
vacuum reduce to three second order ordinary differential equations with one
constraint, thus reducing the dimension of the phase space from infinity to
five.  Even with this comparative simplicity, there has been much debate as to
the existence and nature of chaos in this model.  Most authors have
concentrated on obtaining
Lyapunov exponents (see Eq.~(\ref{Lyap}) and surrounding discussion) of
either the
full equations~\cite{BBE,RJ,HBWS} or discrete
approximations~\cite{Barrow,KLKSS}.  The main difficulty
is that time is a parameter in general relativity, with the same status as
spatial coordinates, so there is no natural variable ``t'' to use
in Eq.~(\ref{Lyap}).
Thus Lyapunov exponents which are nonzero with respect to the discrete time
are zero with respect to the time variable in which the equations take their
simplest form~\cite{BBT}. This difficulty is directly related to what is
known as ``the problem of time'' in quantum cosmology\cite{cosmo}.
Like chaos, quantum mechanics relies on a fixed notion of time to define
evolution.

One approach to circumvent this difficulty has been to represent the dynamics
as geodesic motion in a curved space. If the curvature is negative there is
sensitive dependence on initial conditions and the dynamics may be
chaotic~\cite{SSB,Szylowski}.  Sensitivity to initial conditions is necessary
for chaos, but not sufficient.  Other conditions, such as a compact phase
space and mixing of trajectories are required to prove that a system exhibits
chaos~\cite{Yurtsever}.  This method has the advantage that it
is based on curvature
scalars which do not depend on a choice of coordinate system, however its
applicability to the Mixmaster system has been questioned~\cite{BT}.  There has
also been a study of the homoclinic loops of a Mixmaster Universe containing
matter, albeit with an unrealistic equation of state~\cite{KDD}.  To summarize,
the study of chaos in the Mixmaster Universe has been hampered by a lack of
suitable coordinate invariant quantities to characterize the dynamics.  In
contrast, a Robertson-Walker model (which is isotropic) containing a scalar
field has been shown to exhibit chaotic behavior by other
methods such as the use of Poincare sections, which contain coordinate
independent information~\cite{CE}.

The other class of general relativistic chaotic system, the one which will
concern us, is the dynamics of particles interacting with a given
gravitational and electromagnetic field configuration.  Examples include
particles in the gravitational and electrostatic fields of two or more
black holes~\cite{C90,C91,CP,DFC},
charged particles in a constant magnetic field interacting with gravitational
waves~\cite{VP}, particles near a black hole interacting with gravitational
waves~\cite{BC} and
particles near a black hole immersed in a constant magnetic field~\cite{KV}.
The chaos in these systems has been investigated using
methods borrowed from nonrelativistic systems:  analysis of the periodic
orbits~\cite{C90,C91,CP},Poincare sections~\cite{C91,VP,KV}, Lyapunov
exponents~\cite{DFC,KV}, fractal dimensions~\cite{DFC},
the Chirikov criterion~\cite{VP}, and the Melnikov method~\cite{BC}.
These methods were generalized to relativistic systems
in a fairly straightforward manner (with the possible exception of Lyapunov
exponents) because all of the above gravitational systems are either static
or static with a small periodic perturbation, so there is a well-defined
global time parameter, and the phase space is in some sense static.
Lyapunov exponents are discussed in more detail below.  For
recent reviews of chaos in general relativity, for both the field
equations and the motion of particles, we refer the reader to~\cite{PROC}.

It would appear that fractals do not fit well with general relativity,
as the former are essentially nondifferentiable, but general relativity
is based on smooth manifolds as models for spacetime.  Nevertheless
there have been attempts to use a nondifferentiable manifold as a model
for spacetime, either as resulting from quantum gravity, or in an attempt
to explain the wavelike properties of elementary particles.  This type of
approach is difficult mathematically, and is far from providing a complete
theory at present, although there are some encouraging results.  A number
of recent articles on this subject may be found in~\cite{CSF}.  In contrast,
we show here that fractal basin boundaries provide a particularly reliable
quantification of chaos in general relativity.

\section{Particles in multi-black-hole spacetimes}
\subsection{Formalism}
In nonrelativistic classical mechanics gravity and electrostatics both have
an inverse square force law, so that in any static distribution with equal
mass and charge distributions (in appropriate units: $4\pi\epsilon_{0}=G=c=1$)
the gravitational and electrostatic forces cancel, and the distribution remains
static.

Remarkably, the same situation holds in general relativity.
Majumdar~\cite{Majumdar} and Papapetrou~\cite{Papapetrou} independently
showed that for the static metric:
\begin{equation}
ds^{2}=-U^{-2}dt^{2}+U^{2}(dx^{2}+dy^{2}+dz^{2})\;\;,\label{metric}
\end{equation}
where $U$ is a function of the spatial coordinates, together with the
electrostatic potential
\begin{equation}
A_{t}=U^{-1}
\end{equation}
the Einstein-Maxwell equations reduce to Laplace's equation
\begin{equation}
\nabla^{2}U(x,y,z)\equiv U_{,xx}+U_{,yy}+U_{,zz}=0\;\;.
\end{equation}
Thus, as in nonrelativistic mechanics, there is a static solution of the
field equations for every solution of Laplace's equation.  Hartle and
Hawking~\cite{HH} gave physical interpretation to these solutions, showing
that if $U$ is of the form
\begin{equation}
U=1+\sum_{i=1}^{N}\frac{M_{i}}{\sqrt{(x-x_{i})^2+(y-y_{i})^2+(z-z_{i})^2}}
\;\;,\label{U}
\end{equation}
the Majumdar-Papapetrou (MP) metric corresponds to a system of black holes
with equal charge and mass $M_{i}>0$ and horizons at $(x_{i},y_{i},z_{i})$.
Note that these points are coordinate singularities, that is, a single point
in the MP coordinate system corresponds to a black hole horizon of finite
proper area.  Hartle and Hawking extended the coordinate system to include
the region inside the black holes.

They also showed that for any other form of the solution containing points at
which $U$ is infinite or zero, the singularity is real, and not a result of the
coordinate system chosen.  It is straightforward to use the MP metric to write
down
spacetime solutions corresponding to charged cosmic strings (of finite extent)
or even fractal distributions of mass and charge~\cite{DF}.  A knowledge
of these solutions enriches our understanding of the singularity structure of
more generic solutions of the Einstein-Maxwell equations, although the status
of naked singularities as realistic models of physical objects is doubtful at
present~\cite{Clarke}.

Recently the MP spacetimes have been generalized to include a positive
cosmological constant~\cite{KT,BHKT}.  These solutions describe
coalescing extremal black holes in a de Sitter type universe.  A possible
generalization of this paper would be to investigate the
motion of particles in these spacetimes.  The main differences would be that
the Ricci curvature is negative (leading to instability) and the spacetime
is no longer static.

The general relativistic equation of motion for a particle of charge $e$ and
mass $m$ in combined electromagnetic and gravitational fields is most simply
derived from the super-Hamiltonian~\cite{MTW}
\begin{equation}
{\cal H}=\frac{1}{2}g^{\mu\nu}(\pi_{\mu}-eA_{\mu})(\pi_{\nu}-eA_{\nu})\;\;,
\label{SH}
\end{equation}
where Hamilton's equations
\begin{equation}
\frac{\partial\cal H}{\partial \pi_{\mu}}=\frac{dx^{\mu}}{d\lambda}\;\;,
\qquad
\frac{\partial\cal H}{\partial \mbox{$x$}^{\mu}}=-\frac{d\pi_{\mu}}{d\lambda}
\;\;,
\end{equation}
are written in terms of an affine parameter $\lambda$ related to the particle
proper time by $\tau=m\lambda$.  The mass shell constraint is
\begin{equation}
{\cal H}=-\frac{m^2}{2}\;\;.
\end{equation}
The first Hamilton equation relates the canonical momenta $\pi_{\mu}$ to the
four-velocities:
\begin{equation}
g^{\mu\nu}(\pi_{\mu}-eA_{\mu})=g^{\mu\nu}p_{\mu}=mu^{\nu}\;\;.
\end{equation}
The second equation gives a generalized Lorentz force
equation.  In the MP metric the equations for the spatial components of
the momentum reduce to
\begin{equation}
\frac{dp_{i}}{d\lambda}=\frac{\partial U}{\partial x^{i}}
\left(Up_{0}^{2}+ep_{0}+U^{-3}\sum_{j}p_{j}^{2}\right)\;\;.
\end{equation}
The equations may be simplified slightly by using the four-velocity in an
orthonormal (as opposed to a coordinate) basis
\begin{equation}
u^{\hat{0}}=-\frac{Up_{0}}{m}\;\;,
\qquad u^{\hat{\imath}}=\frac{p_{i}}{Um}\;\;,
\end{equation}
which has a straightforward physical interpretation: the components are
simply $(\gamma,\gamma{\bf v})$ as measured by an observer stationary with
respect to the spacetime.  Writing the components $u^{\hat{\imath}}$ in
vector notation as simply $\bf u$ and $u^{\hat{0}}$ as $\gamma$, and using a
dot for derivatives with
respect to proper time $\tau$, the equations become
\begin{eqnarray}
\dot{\bf u}&=&U^{-2}[(\gamma^{2}+u^{2}-e\gamma/m)\nabla U
-{\bf uu}\cdot\nabla U]\;\;,\\
\dot{\bf x}&=&U^{-1}{\bf u}\;\;,\\
\dot{t}&=&U\gamma\;\;,\\
\gamma&=&\sqrt{1+u^{2}}\;\;.
\end{eqnarray}
These equations are used in the numerical integration, described below.
Because the equations are time-independent, there is a conserved energy,
given by
\begin{equation}
E=-\pi_{0}=U^{-1}(m\gamma-e)\;\;.
\end{equation}
The energy of the particle at infinity is given by $E+e$, rather than simply
$E$, in order to account for the non-zero electromagnetic potential energy
at infinity.  Constancy of the energy is a useful check of the numerical
results, since it is not enforced directly when evolving a trajectory.

\subsection{Hamilton-Jacobi method}
Consider the two-black-hole system, described by the above equations of
motion with the appropriate expression for $U$ from Eq.~(\ref{U}).
Without loss of generality masses $M_{1}$ and $M_{2}$ are placed at $(0,0,1)$
and $(0,0,-1)$ respectively.  We want
to elucidate the structure of the dynamics and classify the problem as
integrable or chaotic, and if it is chaotic, to understand the role of
relativity in quantifying the chaos.

The Newtonian version of the two-black-hole system, that is, a particle
moving in the field of two fixed masses, is integrable, and was first
solved by Euler.  It is instructive to see the effect of relativistic
terms in this solution.  The technique we will use is the Hamilton-Jacobi
method  (chapter 10 of~\cite{Goldstein}), used by Carter to solve the equations
for a
particle moving in the field of a single rotating black hole~\cite{Carter}.
If the Hamilton-Jacobi equation for a particular problem is separable,
the system is integrable (non-chaotic), and all of the constants of motion are
obtained.  Note that our previous short paper~\cite{DFC} uses different
conventions for $E$, $\lambda$, $\cal H$ and the form of the Hamilton-Jacobi
equation than those given below.  Here, the conventions conform to the most
common usage in the literature.

We begin by writing the Hamiltonian using the most natural coordinates
for the system, which in this case are prolate spheroidal coordinates,
used in previous studies~\cite{C90,C91,CP,DFC,Chandra}:
\begin{equation}
x=\sinh\psi\sin\theta\cos\phi\;\;,\qquad
y=\sinh\psi\sin\theta\sin\phi\;\;,\qquad
z=\cosh\psi\cos\theta\;\;.
\end{equation}
The metric~(\ref{metric}) becomes
\begin{equation}
ds^{2}=-U^{-2}dt^{2}+U^{2}\left(Qd\psi^{2}+Qd\theta^{2}+
\sinh^{2}\psi\sin^{2}\theta d\phi^{2}\right)\;\;,
\end{equation}
where
\begin{eqnarray}
U=1+W/Q\;\;,\qquad Q=\sinh^{2}\psi+\sin^{2}\theta\;\;,\nonumber\\
W=(M_{1}+M_{2})\cosh\psi+(M_{1}-M_{2})\cos\theta\;\;.
\end{eqnarray}

The Hamilton-Jacobi equation is the partial differential equation for
$S(x^{\mu},\lambda)$ obtained by taking the equation
${\cal H}=-\partial S/\partial\lambda$
and replacing the momenta $\pi_{\mu}$ in $\cal H$ by
$\partial S/\partial x^{\mu}\equiv S_{,\mu}$, that is,
\begin{equation}
-S_{,\lambda}=-\frac{U^{2}}{2}
\left(S_{,t}-\frac{e}{U}\right)^{2}
+\frac{1}{2U^{2}Q}\left(S_{,\psi}^{2}+S_{,\theta}^{2}\right)
+\frac{S_{,\phi}^{2}}{2U^{2}\sinh^{2}\psi\sin^{2}\theta}\;\;.
\end{equation}
The equation is solved by separation of variables, starting from an ansatz
of the form
\begin{equation}
S=\Lambda(\lambda)+T(t)+\Psi(\psi)+\Theta(\theta)+\Phi(\phi)\;\;.
\end{equation}
Because $\lambda$, $t$, and $\phi$ are cyclic coordinates which do not appear
in $\cal H$, there are three obvious constants of the motion, which give
\begin{equation}
\Lambda=\frac{m^{2}\lambda}{2}\;\;,\qquad
T=-Et\;\;,\qquad \Phi=L_{z}\phi\;\;.
\end{equation}
Here, $L_{z}$ is the $z$ component of angular momentum.  For the full two
black hole
geometry, the remaining equation does not separate, so no further constants
of the motion may be found.  However, in the weak field approximation,
that is, to first order in $W/Q$, we substitute
\begin{equation}
QU^{n}\approx Q+nW\;\;,
\end{equation}
in the Hamilton-Jacobi equation, and it separates to obtain
\begin{eqnarray}
\Psi&=&\int\left\{[4E^{2}+6eE+2e^{2}-2m^{2}](M_{1}+M_{2})\cosh\psi\right.
\nonumber\\& &\left.+[(E+e)^{2}-m^{2}]\sinh^{2}\psi
-L_{z}^{2}/\sinh^{2}\psi+\alpha\right\}^{1/2}d\psi\;\;,\\
\Theta&=&\int\left\{[4E^{2}+6eE+2e^{2}-2m^{2}](M_{1}+M_{2})\cos\theta\right.
\nonumber\\& &\left.+[(E+e)^{2}-m^{2}]\sin^{2}\theta
-L_{z}^{2}/\sin^{2}\theta-\alpha\right\}^{1/2}d\theta\;\;.
\end{eqnarray}
This shows that the weak field approximation is integrable, with the final
constant of the motion being the separation constant $\alpha$.  There is a
clear physical interpretation of this result: The non-relativistic problem
is separable in prolate spheroidal coordinates because the ellipses and
hyperbolae which constitute the lines of constant $\psi$ and $\theta$ are
particle trajectories.  But when the potential becomes of order unity,
relativistic effects become important, including the well-known result
that elliptic orbits precess.  Note that we have not made any approximation
about the velocity of the particle.  A relativistic particle moving in
a weak potential is only slightly deflected, and still does not exhibit chaos.

The above result says very little about the fully relativistic two-centre
problem, except that
prolate spheroidal coordinates are not uniquely suited to studying
this system, except from the point of view of the nonrelativistic limit.
To determine whether the MP problem is chaotic, we must define
parameters to quantify the chaos which arise naturally in the
formalism of general relativity.  The parameters we will use are fractal
dimensions and Lyapunov exponents, evaluated numerically by integrating the
equations of motion.  It may be possible to prove the existence of chaos
analytically, however there is no general method for doing this in a
given dynamical system~\cite{CDF}.  A recent preprint~\cite{Yurtsever} uses
coordinate invariant criteria to study the motion of photons in MP spacetimes,
and concludes that all periodic orbits are unstable, but needs numerical
results such as those given here to back up an argument that this system is
chaotic.

Assuming that this system is chaotic, for which we give very strong numerical
evidence below, the transition to chaos from the integrable Newtonian two
center problem could then be investigated by writing the Hamiltonian as an
integrable, weak field term and a small relativistic perturbation.  As the
strength of the perturbation is increased, the KAM tori of the integrable
system are destroyed by phase space resonances~\cite{BC,AA} leading to
stochastic layers, cantori, and so on.  We use a more direct numerical
approach here, and leave such an investigation to a future paper.

For three or more masses, the Newtonian system is chaotic, except in the
trivial case of test particles with $e=m$, which experience no force at all.
Relativistically, these particles do experience a force if they are moving,
which is proportional to $v^2$ if $v\ll c$, so the dynamics is not trivial.

\section{Qualitative Features}
The equations of motion may be integrated using a 4th order Runge-Kutta
routine with adaptive step size, similar to the one given in~\cite{PFTV}.
The accuracy of the integration may be
checked by a number of methods, including varying the step size controlling
parameter, checking that energy is conserved, and making sure that the
trajectories are physically reasonable.

The results in this and the following section are almost entirely for the two
black hole problem, with masses of $1/3$, placed at $x=z=0$, $y=\pm1$,
and a test particle of zero
charge.  A typical trajectory for this system is shown in Fig.~\ref{0.249}.
Note that, in contrast to the equivalent Newtonian system, there is a finite
cross section for capture by one of the black holes.  The numerical
integration tests for this by stopping whenever the step size becomes smaller
than a predefined limit.

A trajectory with similar initial conditions to the above trajectory is
shown in Fig.~\ref{0.25}.  The outcome is quite different: the particle ends
up in the other black hole.  These trajectories are not particularly
special, but the outcome depends sensitively on the initial conditions.
This characteristic of the dynamics is one of the effects of chaos.

Not all trajectories end up in one of the black holes.  Even for trajectories
with negative energy, which cannot escape to infinity, the particle may orbit
indefinitely.  An example of this is shown in Fig.~\ref{0.538}.  The
phase space may thus be divided into three sets, the basin of attraction of
the black hole at $y=1$, which is marked in black in the following figures,
the basin of attraction of the black hole at $y=-1$, which is marked in
white, and those trajectories which continue to infinite proper time, which
are marked in grey.  Numerically, it is only possible to integrate to a
finite proper time, however it is found that trajectories which fall into
one of the black holes do so within a few orbits.  That is, varying the
maximum time cutoff has virtually no effect on the results, after about
$1000$ (proper) time units.

Some ${\bf u}=0$ slices of phase space are shown in
Figs.~\ref{fig1}---\ref{fig4}.
It is seen that the basins of attraction are quite complicated, and their
mutual boundary appears to be a fractal.  This is another indication that
this system is chaotic.  We will quantify the fractal nature of this
section of the basin boundary in the next section by obtaining a numerical
estimate of its dimension.

Fig.~\ref{fig4} contains a grey region.  As noted above it is not possible
to prove that the trajectories in this region survive for infinite time,
however the picture does not change much if the maximum time is increased,
so this looks to be a real effect.  A typical trajectory in this region is
shown in Fig.~\ref{duality}.  The particle starts at rest at one end of the
trajectory, moves to the other end, and then retraces its steps.  This
orbit is stable, otherwise numerical errors would cause it to deviate
after a few iterations.  It does not seem to have been noted in an earlier
study of the periodic orbits of this system~\cite{CP}.

There are a number of parameters which can be modified in the above system.
If the charge/mass ratio ($e/m$) is equal to one, there is still a force on
a moving test particle, but not on a stationary test particle.  Thus the
${\bf u}=0$ section of phase space is all ``grey'', as none of the particles
ever fall into one of the black holes.  If it is slightly less than one,
however, the result is remarkably similar to that of zero charge, as is
evidenced in Fig.~\ref{0.99}.  With the exception of a slight change of scale
and location
in phase space, this picture is almost indistinguishable from the equivalent
zero charge picture, Fig.~\ref{fig2}.  This is an example of a form of
universality in chaos, where the structure depends only on the topology of
orbits in phase space, and not on the detailed form of the equations.

This also means that there are sharp transitions as parameters are varied.
For example, if, for $e=0$, the mass of the black holes is reduced, thus
moving towards the nonrelativistic limit, the fractal basins gradually take
up more and more of phase space, as the ``capture'' cross section decreases,
until about $M=0.00602$.  At that point, grey regions begin to appear, which
eventually take up the whole of phase space, and become the stable orbits
which characterize all but a set of zero measure of Newtonian phase space.
The phase space for $M=0.006$ and a typical ``Newtonian-like'' stable orbit
are shown in Figs.~\ref{m60},~\ref{m60traj}.  The line for this trajectory
is thick because the particle does not follow exactly the same path for each
orbit.  The is also true of the trajectory shown in Fig.~\ref{duality},
but to a smaller degree.

Also, we may vary the initial velocity.  Figs.~\ref{v0.4}--\ref{v1.5}
show a plot of
the $M=1/3$, $e=0$ phase space with an initial velocity which is constant
over the plot.  Note that the boundaries of the grey regions appear to be
smooth curves.  This is partly explained by the fact that the region in which
the total energy is positive is roughly hour-glass shaped.  The
boundary between the basins of attraction still appears to be a fractal.

Finally, we may consider three-black-holes, as shown in Figs.~\ref{3H1}--%
\ref{3H5}.  Here, black, white and grey correspond to the basins of attraction
of the three holes, which are placed at the vertices of an equilateral
triangle of side length $\sqrt{3}$.  Each hole has mass $1/3$.  This section
of phase
space has the curious feature that between each black and white region there
is a grey region, and so on.  Or, to put it another way, there is only one
basin boundary, and that boundary is the boundary of all three regions.  This
property is similar to the Julia sets generated by applying Newton's method
to cubic equations~\cite{Falconer}, and is only possible if the boundary is a
fractal.  A proof of this property involves an investigation of orbits
which just escape falling into a particular black hole.  These circle the
black hole an infinite number of times, and are arbitrarily close to orbits
which circle a large number of times but then escape and head towards either
of the other holes.  Thus the boundary is arbitrarily close to all three
basins of attraction.  The Newtonian limit of this system is chaotic, but does
not have attractors.  A similar nonrelativistic system with attractors and
boundary basins is the magnetic pendulum~\cite{MGOY}.

\section{Measures of Chaos}
\subsection{Fractal Dimensions}
There are a number of different fractal dimensions used in the
literature~\cite{Falconer,Mandelbrot}.  The easiest to estimate
numerically is the box dimension, which is a non-negative
real number assigned to a subset $F$ of $E$-dimensional Euclidean space $R^E$.
There are several equivalent
definitions.  The definition used here to estimate the dimension of
the basin boundary numerically is as follows:  Fill a section of $R^E$ with a
grid of
$E$-dimensional cubes of side length $\delta$.  Let $N_{\delta}(F)$ be the
number of grid cubes containing $F$.  Then the box dimension is
\begin{equation}
d_{B}(F)=-\lim_{\delta\rightarrow 0}\frac{\ln N_{\delta}(F)}{\ln \delta}\;\;.
\end{equation}
Other equivalent definitions include covering the set $F$ with a minimum
number of spheres or other shapes or finding the maximum number of disjoint
spheres with centers in $F$.  The box dimension involves a limit, so there
is no guarantee that it is actually defined for a given set $F$.

If the space is non-Euclidean, for
example the curved pseudo-Riemannian spacetime of general relativity, then
the concept of a ``cube of size $\delta$'' in the definition of dimension
becomes ill-defined.  However
it is not necessary to cover the actual fractal object with cubes: Choose
a coordinate system, and calculate the box dimension of the object in
coordinate space.  An important result is that any diffeomorphism,
that is, differentiable coordinate transformation, leaves the box dimension
invariant, so the original choice of coordinate system was irrelevant,
and the box dimension of the object is well defined.  Thus we can ignore
the curvature of the MP spacetime, and simply calculate the dimension of
the basin boundary as if it were embedded in a flat Euclidean space.

The value of $d_{B}$ was evaluated using the above equation for a region
in phase space near that shown in Fig.~\ref{fig2}.  The region, containing
$2520^{2}$ points, was covered by a grid.  Each square of size $\delta$
(a factor of $2520$) was counted if it contained points of different colour,
that is, trajectories with different final outcomes.  Fig.~\ref{dimplot}
shows a plot
of $\ln N$ vs $\ln\epsilon$.  The straight line is a least squares fit to all
but the
three smallest and ten largest values of $\delta$, and gives a dimension of
$1.43$.  The uncertainty is about $0.03$, which is typical for such
investigations~\cite{Obert}.  This uncertainty arises from the sampling of
regions in phase space and ambiguity in how to actually perform the fit.
The curvature of the graph in Fig.~\ref{dimplot}
towards small $\delta$ is
well explained by the fact that for small box sizes, it is possible to
miss a box which actually contains part of the boundary, and hence
underestimate $N_{\delta}$.

The presence of fractal boundary basins indicates that there are
non-differentiable (non-smooth) structures in phase space, which implies the
system is chaotic. In contrast, an integrable Hamiltonian system has enough
constants of motion to determine the motion completely, and these constants
must be smooth functions of the phase space coordinates. The usual provisos
apply: The numerical methods give estimates of the fractal dimension over a
range of scales, but can never take exact $\delta\rightarrow 0$ limits.
Nevertheless the statement that the boundary is a fractal is quite convincing.

We have performed the same analysis to Fig.~\ref{3H5} for the three-black-hole
problem.  In this case there are $1260^2$ points, and the resulting graph,
leading to a dimension of $1.47$, is shown in Fig.~\ref{dimplot2}.

\subsection{Lyapunov exponents}
Another indicator of chaos for nonrelativistic systems is the presence of
positive Lyapunov exponents (defined below) for a non-isolated set of
trajectories.
The latter condition is necessary because unstable fixed points generally have
positive Lyapunov exponents, and occur in integrable as well as chaotic
systems.  The difference is that the unstable trajectories are isolated in
integrable systems, whereas chaotic systems may be composed entirely of
unstable trajectories.  Typically chaotic systems contain both stable and
unstable regions in phase space, with a fractal boundary between the two.

The Lyapunov exponents $\lambda_{k}$ in flat spacetime are defined by choosing
a point $x$ in
phase space, at the center of a ball of radius $\epsilon\ll 1$.  After a time
$t$ the ball evolves into an ellipsoid with semi-axes $\epsilon_{k}(t)$,
where $k$ ranges from one to the dimension of the phase space.  The
Lyapunov exponents are
\begin{equation}
\lambda_{k}(x)=\lim_{t\rightarrow\infty}\lim_{\epsilon\rightarrow 0}
\frac{1}{t}\ln\frac{\epsilon_{k}(t)}{\epsilon}\;\;,\label{Lyap}
\end{equation}
assuming the limits exist.  The $\lambda_{k}$ are constant along a trajectory,
and are often constant over larger regions of phase space, such as the basin
of an attractor.

There are a number of subtleties associated with the definition of Lyapunov
exponents in curved spaces, and in particular the MP spacetimes.  What time
parameter should be used for $t$?  This question has particularly plagued
the Mixmaster problem, as discussed in Section~II.  In a general
spacetime the only time parameter of any special significance for a trajectory
is the proper time $\tau$.  If this is used in Eq.~(\ref{Lyap}) the result is
a measure of the local instability of phase space trajectories, but does not
give information about the global properties of the system.  The MP spacetime
is static, so that there are a set of distinguished observers ``at rest''
with respect to the black holes.  The time as measured by these observers
depends on their position due to the gravitational redshift, but at infinity
approaches a constant rate, given by $t$, which appears in the metric,
Eq.~(\ref{metric}).  We will use this parameter, following Ref.~\cite{KV}.

The other difficulty is that it is not obvious how to calculate distances in
phase space, given that the original spacetime is curved.  Here, as with the
fractal dimension, any metric gives the same answers, provided that the
trajectory returns arbitrarily close to the starting point, as this causes the
metric terms to cancel in the expression $\epsilon_{k}(t)/\epsilon$.  The
above
condition is not very restrictive, as it applies to any trajectories remaining
in a compact region of the phase space, including periodic orbits.  In our
case, this means that Lyapunov exponents are well defined for trajectories
which do not fall into a black hole or escape to infinity.  In the former
case, the limit $t\rightarrow\infty$ is also not sensible, while in the
latter, the exponents are
zero for any metric in phase space which reduces to the special relativistic
one in the flat spacetime limit.

The numerical method by which Lyapunov exponents are calculated is described
in Ref.~\cite{SN}.  For a
general set of coupled ODE's
\begin{equation}
\dot{x}^{i}=f^{i}({\bf x})\;\;,
\end{equation}
the equation for a perturbation $\delta x^{i}$ is
\begin{equation}
\delta\dot{x}^{i}=\delta x^{j}\frac{\partial}{\partial x^{j}}f^{i}({\bf x})
\;\;,\label{linear}
\end{equation}
which is a linear equation containing a known function of $\bf x$, which is
unknown.  An orthonormal basis of perturbations is integrated at the same
time as the equations of motion, and a Gram-Schmidt orthonormalization is
carried out periodically to ensure that one vector is lined up in the fastest
growing direction, one in the second, and so on; this also ensures that
the exponentially growing solutions do not generate overflow or roundoff
errors.  The Lyapunov exponents are obtained simply by adding the logarithm
of the scaling factors in the Gram-Schmidt algorithm, and dividing by the
total time.  We call this the ``straight'' algorithm.

In our problem we have the additional difficulty that, in certain regions of
phase space, such as those shown in the figures in the previous section,
almost all of the
trajectories fall into one of the black holes, yet those that do not still
have meaningful Lyapunov exponents which we wish to estimate.  In an attempt
to get around this problem we have used an algorithm which periodically
checks whether the trajectory survives a specified time.  If it does not,
the position in phase space is shifted randomly by a small amount.  We have
checked that this does not cause the energy to drift perceptibly. The shifting
algorithm is difficult
to use in that there are a number of parameters to choose, and the shifting
routine may fail to find any suitable trajectories.  In fact the trajectory
given in the table below is the only one we have tried for which it has
succeeded for a
long integration time.  Of course, it gives the same result as the straight
algorithm if shifting is not necessary.

It is sometimes possible to calculate Lyapunov exponents analytically.
The equations of motion are known, so if an analytic solution for a trajectory
can be found, the linear Eq.~(\ref{linear}) can be solved, giving
a matrix whose eigenvalues are the Lyapunov exponents.  One such trajectory
is the unstable fixed point midway between the black holes.

For the general
case of a fixed point, $U_{,i}=0$, where $i$ indicates $x$ or $y$, and the
comma is a partial derivative, as usual.  The linearized equations become
\begin{equation}
\frac{d}{d\tau}\left(\begin{array}{c}\delta x\\\delta y\\\delta u_{x}\\
\delta u_{y}\end{array}\right)=\left(\begin{array}{cccc}0&0&U^{-1}&0\\
0&0&0&U^{-1}\\aU_{,xx}&aU_{,xy}&0&0\\aU_{,xy}&aU_{,yy}&0&0\end{array}\right)
\left(\begin{array}{c}\delta x\\\delta y\\\delta u_{x}\\
\delta u_{y}\end{array}\right)\;\;,
\end{equation}
with
\begin{equation}
a=\frac{1}{U^{2}}\left(1-\frac{e}{m}\right)\;\;,
\end{equation}
and have solutions of the form $e^{l\tau}$ where $l$ is an eigenvalue of the
above matrix, that is,
\begin{equation}
l=\pm\sqrt{\frac{1-e/m}{2U^{3}}}\sqrt{U_{,xx}+U_{,yy}
\pm\sqrt{(U_{,xx}-U_{,yy})^{2}+4U_{,xy}^{2}}}\;\;.
\end{equation}

The Lyapunov exponents are closely related to the values of $l$, but differ in
the following ways: An imaginary value of $l$ leads to oscillatory solutions
which have Lyapunov exponent zero, and the exponents are calculated using the
global time $t$, so that the remaining exponents are given by
$\lambda=U^{-1}l$, using the metric Eq.~(\ref{metric}).  Thus the Lyapunov
exponents may be calculated for a fixed point, and depend only on the values
of $U$ and its derivatives at the point, and the value of $e/m$.  The analytic
value for the fixed point in the case of two-black-holes of mass $1/3$ is
given in Table~\ref{Loopy}, and agrees with the numerically evaluated result
to five significant figures.

If $e=m$, the above expression yields zero for the Lyapunov exponents, which
can be understood physically because the force on the particle
is proportional to $v^2$ for small $v$, so the equations linearized about
the stationary solution are trivial.  These analytic values are useful in
checking the numerics,
but do not yield any information about the chaos, since they apply to
trajectories which are not in chaotic regions.

The trajectories which have been integrated numerically in Table~\ref{Loopy}
are the unstable fixed point at the origin, which compares very closely with
the analytic result, the trajectory plotted in Fig.~\ref{0.538}, the
trajectory plotted in Fig.~\ref{duality}, a trajectory which oscillates up
and down the $x$-axis, and a trajectory ``on'' the boundary which has needed
to be shifted.  All of these have been integrated for $t=10^5$ time units.
The ``algorithm'' is either the analytic calculation given above, the straight
algorithm, or the shifting algorithm.

It is clear that the numerical results reflect the symmetries of the equations.
The sum of the exponents is zero, due to Liouville's theorem.  In addition,
two of the exponents are (approximately) zero due to the one constant of motion
$E$.  As the length of the integration increases, these decrease, showing that
they are due to the finite averages used, and not numerical errors in the
equations.

Two of the orbits appear to have all four exponents zero.  This is expected,
since they are numerically stable in that a small perturbation
of the initial conditions does not cause a qualitative difference in the
trajectory.

The trajectory which oscillates along the $x$-axis is in the unusual
position of being exactly calculable (in theory: the result is a complicated
integral), but arbitrarily close to the fractal boundary.  See Fig.~%
\ref{medium}.  It is only a special orbit in that the integration can be
carried on for an indefinite time due to the symmetry of the equations.
Otherwise it is similar to the other unstable periodic orbits of this system.
It might be possible to analyze the Lyapunov exponents of this system by noting
that an arbitrary point on the basin boundary is arbitrarily close to periodic
orbits.  The Lyapunov exponents for the periodic orbits could be evaluated
with some definiteness, because only a finite time is required.

\section{Discussion}
We have seen that there are difficulties in applying the standard tests of
chaos to general relativistic systems.  Because the spacetime and the phase
space derived from it depend on an arbitrary choice of coordinate system,
quantities which depend on distances in phase space, such as Lyapunov
exponents become poorly defined and unsuitable as a test for chaos.  For the
MP problem, the spacetime is static, so a natural global time variable exists,
and these problems become less severe.  In any case, quantities which are
topological in nature, or at least coordinate independent, such as the fractal
nature of structures in phase space, are equally good for relativistic
problems.  Poincare sections also fall into this category.

The motion of particles in systems of two or three fixed black holes exhibits
many of the features common to non-relativistic chaotic systems, such as
sensitive dependence on initial conditions leading to nonzero Lyapunov
exponents and fractal basin boundaries.  There are islands of stability
hidden in the chaos.  The structure of the phase space depends only slightly
on the charge of the particle.  What is unusual about this problem is the
presence of attractors in a non-dissipative system.  The fractal boundaries
of the basins of these attractors are particularly useful to quantify the
chaos, since their dimension does not depend on the chosen coordinate system.

Since the two-black-hole problem has an
integrable Newtonian limit, it should be possible to observe the break up
of the KAM tori explicitly as the mass of the black holes is increased from
zero.  What makes this limit most interesting is that
the relativistic effects lead to capture of the particle, a process that is
not present in the non-relativistic system.

In closing, we remark that fractal geometry is particularly well suited to
studying chaos in the most geometrical of theories --- general relativity.

\begin{figure}
\caption{A trajectory with initial conditions ($x=3$, $y=0.249$,
$u_{x}=0$, $u_{y}=0)$.}
\label{0.249}
\end{figure}

\begin{figure}
\caption{A trajectory with initial conditions ($x=3$, $y=0.25$,
$u_{x}=0$, $u_{y}=0)$.}
\label{0.25}
\end{figure}

\begin{figure}
\caption{A trajectory with initial conditions ($x=3$, $y=0$,
$u_{x}=0$, $u_{y}=0.538)$.}
\label{0.538}
\end{figure}

\begin{figure}
\caption{A ${\bf u}=0$ section of phase space.}
\label{fig1}
\end{figure}

\begin{figure}
\caption{A subset of Fig.~\protect\ref{fig1}}
\label{medium}
\end{figure}

\begin{figure}
\caption{A subset of Fig.~\protect\ref{medium}.}
\label{fig2}
\end{figure}

\begin{figure}
\caption{A subset of Fig.~\protect\ref{fig2}.}
\label{fig3}
\end{figure}

\begin{figure}
\caption{A subset of Fig.~\protect\ref{fig3}.}
\label{fig4}
\end{figure}

\begin{figure}
\caption{A trajectory with initial conditions ($x=2.8458$, $y=0.263$,
$u_{x}=0$, $u_{y}=0)$.}
\label{duality}
\end{figure}

\begin{figure}
\caption{A ${\bf u}=0$ section of phase space with $e=0.99m$.}
\label{0.99}
\end{figure}

\begin{figure}
\caption{A ${\bf u}=0$ section of phase space with $M=0.006$.}
\label{m60}
\end{figure}

\begin{figure}
\caption{A trajectory with $M=0.006$ and initial conditions ($x=4.5$, $y=2$,
$u_{x}=0$, $u_{y}=0)$.}
\label{m60traj}
\end{figure}

\begin{figure}
\caption{A section of phase space with $u_{x}=0$, $u_{y}=0.4$.}
\label{v0.4}
\end{figure}

\begin{figure}
\caption{A section of phase space with $u_{x}=0$, $u_{y}=0.7$.}
\label{v0.7}
\end{figure}

\begin{figure}
\caption{A section of phase space with $u_{x}=0$, $u_{y}=1.5$.}
\label{v1.5}
\end{figure}

\begin{figure}
\caption{A section of three-black-hole phase space with ${\bf u}=0$.}
\label{3H1}
\end{figure}

\begin{figure}
\caption{A subset of Fig.~\protect\ref{3H1}.}
\label{3H4}
\end{figure}

\begin{figure}
\caption{A subset of Fig.~\protect\ref{3H4}.}
\label{3H5}
\end{figure}

\begin{figure}
\caption{The dimension of a section of two-black-hole phase space is $1.43$.}
\label{dimplot}
\end{figure}

\begin{figure}
\caption{The dimension of a section of three-black-hole phase space is $1.47$.}
\label{dimplot2}
\end{figure}

\begin{table}
\caption{Lyapunov Exponents for the Two-Black-Hole Problem}
\label{Loopy}\centering
\begin{tabular}{ccllll}
Initial Conditions&Algorithm&$\lambda_{1}$&$\lambda_{2}$&$\lambda_{3}$&
$\lambda_{4}$\\\hline
(0,0,0,0)&Analytic&$\frac{18}{25\sqrt{5}}$&0&0&$-\frac{18}{25\sqrt{5}}$\\
(0,0,0,0)&Straight&0.32199&0.00001&--0.00001&--0.32199\\
(3,0,0,0.538)&Straight&0.00007&0.00002&--0.00001&--0.00007\\
(2.8458,0.263,0,0)&Straight&0.00005&0.00000&--0.00001&--0.00003\\
(3,0,0,0)&Straight&0.11222&0.00007&--0.00007&--0.11222\\
(3.33467,0.23509,0,0)&Shifting&0.03609&0.00006&--0.00006&--0.03609
\end{tabular}
\end{table}

\end{document}